\documentclass[aps,prd,floatfix,nofootinbib]{revtex4} 
\usepackage{graphics}
\usepackage{graphicx}

\usepackage{latexsym}
\usepackage[cp866]{inputenc}
\usepackage[english]{babel}
\input epsf

\begin{document}

\title{\boldmath Phenomenological description of the $\pi^-\pi^+$
$S$-waves in $D^+\to\pi^-\pi^+\pi^+$ and $D^+_s\to\pi^-\pi^+\pi^+$
decays: The problem of phases.}
\author{N.N. Achasov\footnote{achasov@math.nsc.ru} and G.N.
Shestakov\footnote{shestako@math.nsc.ru}} \affiliation{Laboratory of
Theoretical Physics, S.L. Sobolev Institute for Mathematics, 630090,
Novosibirsk, Russia}


\begin{abstract}
We present a phenomenological description of the LHCb data for the
magnitudes and phases of the $\pi^-\pi^+$ $S$-wave amplitudes in the
$D^+\to\pi^-\pi^+\pi^+$ and $D^+_s\to\pi^-\pi^+\pi^+$ decays. We
operate within a simple model that takes into account the known pair
interactions of particles in coupled channels. The seed complex
amplitudes for various intermediate state production are assumed to
be independent of the energy; their values are determined by
fitting. This model gives a satisfactory description of virtually
all features of the energy dependence of the experimentally measured
$S$-wave amplitudes in the $D^+\to\pi^-\pi^+ \pi^+$ and $D^+_s\to
\pi^-\pi^+\pi^+$ decays in the regions
$2m_\pi<m_{\pi^-\pi^+}<1.39\mbox{ GeV}$ and $2m_\pi<m_{\pi^-\pi^+}
<1.29\mbox{ GeV}$, respectively.
\end{abstract}

\maketitle

\section{Introduction}


Measurements of three-body decays of $D$- and $D_s$-mesons into
$\pi^-\pi^+\pi^+$, $K^-\pi^+\pi^+$, $K ^+K^-\pi^+$, $K^-K^+K^+$,
etc. \cite{Aa22,Ai01a,Ai01,Li04,Bo07,Bo08,Au09,Sa11,No15,Re16,
Aa19,Ab21,Aa22a,Ze22,PDG22} represent the most important extension
of the classical studies of three-pion decays of strange mesons
$K\to\pi\pi\pi$ \cite{Am22,PDG22} into a family of charmed
pseudoscalar states. Information about the resonant structures in
the two-body mass spectra in these decays is obtained from the
Dalitz plot fits using the isobar model \cite{Aa22,Ai01a,Ai01,Li04,
Bo07,Bo08,Au09,Sa11,No15,Re16,Aa19,Ab21,Aa22a,Ze22,PDG22} and
quasimodel-independent partial wave analysis \cite{Aa22,Aa22a,
Ai01,Bo08,Au09,Ab21,Re16}. Further, we will speak  about the
$D^+\to\pi^-\pi^+\pi^+$ and $D^+_s \to\pi^-\pi^+\pi^+$ decays for
which the LHCb Collaboration has recently obtained the detailed
high-statistics data \cite{Aa22, Aa22a}. For the data analysis, the
amplitude of the $D^+\to\pi^-\pi^+\pi^+$ decay \cite{Aa22}  was
approximated by the coherent  sum (symmetrized with respect to the
permutation of two identical pions) of the $S$-wave contribution and
higher-spin waves (the same approximation was also used for the
amplitude of the $D^+_s\to\pi^-\pi^+\pi^+$ decay \cite{Aa22a}),
\begin{eqnarray}\label{Eq1}\mathcal{A}(s_{12},s_{13})=\left[
\mathcal{A}_{S{\scriptsize\mbox{-wave}}}(s_{12})+\sum_i a_i
e^{i\delta_i}\mathcal{A}_i(s_{12},s_{13})\right]+(s_{12}
\leftrightarrow s_{13}),\end{eqnarray} where $s_{12}=(p_1+p_2)^2$
and $s_{13}=(p_1+p_3)^2$ are the squares of the invariant masses of
two different $\pi^-\pi^+$ pairs ($\pi^-_1\pi^+_2$ and
$\pi^-_1\pi^+_3$); $p_1$, $p_2$, $p_3$ are the four-momenta of the
final pions. The first term in square brackets is the $S$-wave
amplitude,
\begin{eqnarray}\label{Eq2}\mathcal{A}_{S{\scriptsize\mbox{-wave}
}}(s_{12})= a_0(s_{12})e^{i\delta_0(s_{12})}.\end{eqnarray} The
values of the real functions $a_0(s_{12})$ and $\delta_0(s_{12} )$
were obtained by the Dalitz plot fitting for 50 intervals (knots)
into which the accessible region of $\sqrt{s_{1 2}}\equiv m_{\pi^-
\pi^+}$ ($2m_\pi<m_{\pi^-\pi^+}<m_{D(D_s)}-m_ \pi$) was divided
\cite{Aa22, Aa22a}. This technique allows one to obtain information
about the $\pi^-\pi^+$ $S$-waves in the $D^+\to\pi^-\pi^+\pi^+$ and
$D^+_s\to\pi^-\pi^+\pi^+$ decays without any model assumptions about
their composition [i.e., about the contributions of the states
$f_0(500)$, $f_0(980)$, $f_0(1370)$, $f_0(1500)$, etc.]. The
motivation for applying this method is the presence of overlapping
wide and narrow light scalar resonances in the region below 2 GeV
with poorly-known masses and widths. The LHCb data on the $S$-wave
amplitudes in the $D^+\to\pi^-\pi^+\pi^+$ \cite{Aa22} and $D^+_s\to
\pi^-\pi^+\pi^+$ \cite{Aa22a} decays are shown below in Figs. 3 and
4. The $S$-wave contributions in these decays are dominant. They
account for approximately 62\% and 85\% of the full decay rate of
$D^+$ and $D^+_s$ into $\pi^-\pi^+\pi^+$, respectively. In turn, the
amplitudes of the $P$- and $D$-waves, represented by the terms in
the sum in Eq. (\ref{Eq1}), were approximated in the isobar model by
the contributions of the known resonances $\rho^0(770)$,
$\omega(782)$, $\rho^0(1450)$, $\rho^0(170 0)$, $f_2(1270)$, and
$f'_2(1525)$. The amplitude $\mathcal{A}_i(s_{12},s_{13})$ of
resonance $R_i$ includes the Breit-Wigner complex resonant
amplitude, angular distribution, and Blatt-Weiskopf barrier factors
(for more details of the parametrization see Refs. \cite{Aa22,
Aa22a}). The magnitude and phase of the $R_i$ production amplitude,
$a_i$ and $\delta_i$, are free (independent of $s_{12}$ and
$s_{13}$) parameters within the isobar model. Their values relative
to the magnitude and phase of the amplitude of the selected
reference subprocess (which are taken to be 1 and 0$^\circ$,
respectively) were also determined in Refs. \cite{Aa22,Aa22a} from
the fits to the data.

The data on the values and energy dependence of the phases of the
$S$-waves in the $\pi^-\pi^+$ channel obtained from the $D^+\to\pi^-
\pi^+\pi^+$ and $D^+_s\to\pi^-\pi^+\pi^+$ decays and $\pi^+\pi^-\to
\pi^+\pi^-$ reaction are discussed in detail and compared with each
other in Ref. \cite{Aa22a}. Obvious differences between all three
phases indicate deviations from the Watson final-state interaction
theorem \cite{Wa52} in the $D^+\to\pi^-\pi^+\pi^+$ and $D^+_s\to
\pi^-\pi^+\pi^+$ decays. This fact is also evidence of the important
role of intermediate multibody hadronic interactions (multiquark
fluctuations) on the formation of the phases of  the production
amplitudes of final two-body subsystems in these and related decays
(for example, in $D^+\to K^-\pi^+\pi^+$) \cite{Aa22a,Ro11,Fr14,
Ro15,No15,Re16,Lo16,Wa22}. In general, the problem of explaining the
specific values of the  phases $\delta_i$ included in Eq.
(\ref{Eq1}) and the energy dependence  of the $S$-wave phases
$\delta_0(s_{12})$ seems to be key for elucidation of the mechanisms
of the $D^+\to\pi^-\pi^+\pi^+$ and $D^+_s\to\pi^- \pi^+\pi^+ $
decays.

This paper presents a phenomenological description of the LHCb data
for the magnitudes and phases of the $S$-wave amplitudes of the
$\pi^-\pi^+$ systems produced in the $D^+\to\pi^-\pi^+\pi^+$ and
$D^+_s\to\pi^-\pi^+\pi^+$ decays. Our model is described in Sec. II.
The fittings to the data on $S$ waves in the decays of $D$ and $D_s$
mesons are presented in Secs. III and IV, respectively. Predictions
for the $\pi^0\pi^0$ $S$-waves in the $D^+\to\pi^+\pi^0\pi^0$ and
$D^+_s\to\pi^+\pi^0\pi^0$ decays are made in Sec. V. The results of
our analysis are briefly formulated in Sec. VI.


\section{A phenomenological model for the $S$-waves}


As is well known,  light scalar mesons are richly produced in the
reactions $\pi^+\pi^-\to\pi^+\pi^-$ and $\pi^+\pi^-\to K\bar K$,
information about which is extracted from the more complicated
peripheral processes $\pi^\pm N\to[(\pi\pi),(K\bar K)](N,\Delta)$
dominated by the one-pion exchange mechanism. We will assume that in
the processes in which the initial state is not the $\pi\pi$
scattering state, the light scalar mesons $f_0(500)$ and $f_0(980)$
are produced in interactions of intermediate pseudoscalar mesons
$\pi^+$ with $\pi^-$, $\pi^0$ with $\pi^0$, and $K$ with $\bar K$.
Note that such a mechanism is quite consistent with the hypothesis
of a four-quark ($q^2\bar q^2$) nature of light scalars
\cite{Ja77,Ac89, Ac98,Ac03a}. The scheme of their formation in the
$D^+\to\pi^-\pi^+\pi^+$ and $D^+_s\to\pi^- \pi^+\pi^+$ decays is
graphically represented in Fig. 1. At the first step, the valence
$c$-quark decays into light quarks, the initial states of the
$D^+=c\bar d$ and $D^+_s=c\bar s$ mesons ``boil up'', passing into a
mixture of various quark-gluon fluctuations, which are then combined
into pions, kaons, etc. The latter can additionally enter into pair
interactions with each other in the final state. We take into
account the seed three-body $S$-wave fluctuations
$D^+/D^+_s\to\pi^+\pi^+\pi^-$, $D^+/D^+_s\to\pi^+\pi^0\pi^0$,
$D^+/D^+_s\to\pi^+K^+K^-$, and $D^+/D^+_s\to\pi^+ K^0\bar K^0$ (the
corresponding amplitudes are shown in Fig. 1 by thick black dots).
In so doing, the $f_0(500)-f_0(980) $ resonance complex is produced
as a result of $\pi\pi$ and $K\bar K$ interactions in the final
state. The amplitudes corresponding to these subprocesses are
indicated in Fig. 1 as $T_{ab\to \pi^+\pi^-}$, where $ab=\pi^+
\pi^-$, $\pi^0\pi^0$, $K^+K^-$, $K^0\bar K^0$.

\begin{figure}  [!ht] 
\begin{center}\includegraphics[width=14cm]{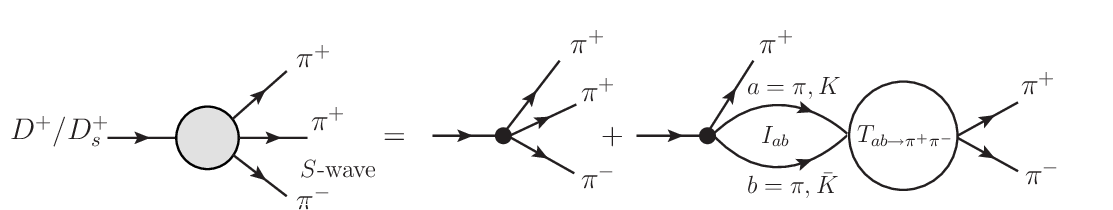}
\caption{\label{Fig1} The $f_0(500)-f_0(980)$ resonance complex
production amplitude in the $D^+\to\pi^-\pi^+\pi^+$ and
$D^+_s\to\pi^- \pi^+\pi^+$ decays. Contributions of the intermediate
states $ab=\pi^+\pi^-$, $\pi^0\pi^0$, $K^+K^-$, $K^0\bar K^0$ are
summed. }\end{center}\end{figure}

According to this figure, we write the $S$-wave amplitude
$\mathcal{A}_{S{\scriptsize\mbox{-wave}}}(s_{12})= a_0
(s_{12})e^{i\delta_0 (s_{12})}$ for the $D^+/D^+_s\to\pi^-
\pi^+\pi^+$ decay (taking into account the renaming $s_{12}\equiv
s\equiv m^2_{\pi^-\pi^+}$) in the following form
\begin{eqnarray}\label{Eq3}
{\mathcal{A}_{S\scriptsize\mbox{-wave}}(s)=a_0(s)e^{i\delta_0(s)
}}=\lambda_{\pi^+\pi^-}+\sum_{ab}\lambda_{ab}I_{ab}(s)\xi_{ab}\,
T_{ab\to\pi^+\pi^-}(s),\end{eqnarray} where $\xi_{ab}=1/2$ for $ab=
\pi^0 \pi^0$ and $=1$ in other cases; $T_{\pi^+\pi^-\to\pi^+ \pi^-}
(s)=\frac{2}{3}T^0_0(s)+\frac{1}{3}T^2_0(s)$, $T_{\pi^0\pi^0 \to
\pi^+\pi^-}(s)=\frac{2}{3}[T^0_0(s)-T^2_0(s)]$, where $T^0_0(s)$ and
$T^2_0(s)$ are the $S$-amplitudes of the reaction $\pi\pi\to\pi\pi$
in the channels with isospin $I=0$ and 2, respectively,
$T^I_0(s)=[\eta^I_0(s)\exp(2i\delta^I_0(s))-1]/(2i\rho_{\pi\pi}(s))$,
where $\eta^I_0(s)$ and $\delta^I_0(s)$ are the corresponding
inelasticity and phase of $\pi\pi$ scattering ($\eta^0_0(s)=1$ at
$s<4m^2_{K^+}$, and $\eta^2_0(s)=1$ in the whole region of $s$ under
consideration), $\rho_{\pi\pi} (s)=\sqrt{1-4m^2_\pi/s}$. For the
$S$-wave transition amplitudes $K\bar K\to\pi\pi$ we have
$T_{K^+K^-\to\pi^+\pi^-}(s)=T_{K^0\bar K^0\to\pi^+ \pi^-}(s)$ and
$T_{K\bar K\to\pi\pi}(s)=T_{\pi\pi\to K\bar K}(s)$. The function
$I_{ab}(s)$ is the amplitude of the $ab$ loop. Above the $ab$
threshold, $I_{ab}(s)$ has the form
\begin{eqnarray}\label{Eq4}
I_{ab}(s)=C_{ab}+\rho_{ab}(s)\left(i-\frac{1}{\pi}\ln\frac{
1+\rho_{ab}(s)}{1-\rho_{ab}(s)}\right),\end{eqnarray} where
$\rho_{ab}(s)=\sqrt{1-4m^2_a/s}$ (we put $ m_{\pi^+}=m_{\pi^0}
\equiv m_{\pi}$ and take into account the mass difference of $K^+$
and $K^0$) if $\sqrt{s}<2m_K$, then $\rho_{K\bar K}(s)\to
i|\rho_{K\bar K}(s)|$, and $C_{ab}$ is a real subtraction constant
in the $ab$ loop. $C_{\pi^+\pi^-}=C_{\pi^0\pi^0}\equiv C_{\pi\pi}$,
$C_{K^+K^-}=C_{K^0\bar K^0}\equiv C_{K\bar K}$, and
$I_{\pi^+\pi^-}(s)=I_{\pi^0\pi^0}(s)\equiv I_{\pi\pi}(s)$.

The  seed $S$-wave amplitudes $\lambda_{ab}$ in Eq. (\ref{Eq3}) are
approximated by complex constants. They are free parameters of the
model along with the constants $C_{ab}$. A similar model approach
has already been applied to the decays $D^+\to\pi^-\pi^+\pi^+$
\cite{Bo07}, $D/D_s\to\pi^+\pi^-e^+\nu_e$ \cite{Ac20}, and
$J/\psi\to\gamma\pi^0\pi^0$ \cite{Ac21}. In fact, we are dealing
with the description of the data on the $S$-wave components of
$D^+/D^+_s \to\pi^-\pi^+\pi^+$ decays in the spirit of the isobar
model in which instead of the resonant Breit-Wigner distributions
one uses the known amplitudes $T^0_0(s)$, $T^2_0(s)$, and $T_{\pi\pi
\to K\bar K}(s)$. All nontrivial dependence on $s$ is introduced
into $\mathcal{A}_{S\scriptsize\mbox{-wave}}(s)$ by these
amplitudes. In their meaning, the absolute values and phases of the
amplitudes $\lambda_{ab}$ in Eq. (3) do not differ from the
amplitudes $a_i$ and phases $\delta_i$ included in Eq. (1). In the
isobar model, all these quantities are considered constant because
they depend only on the total energy of the system, i.e.,
$m_{\pi^-\pi^+\pi^+}= \sqrt{(p_1+p_2+p_3)^2}=M_{D/ D_s}$ in our
case, and do not depend on the subenergy $m_{\pi^-\pi^+}$. In
particular, the imaginary parts of $\lambda_{ab}$ are understood as
a result of the three-body final state interactions dressed the $c$
quark weak-decay vertices. Their presence due to the real and
quasireal intermediate states that can appear at $m_{\pi^-\pi^+
\pi^+}=M_{D/D_s}$ in the input channel. It is the complex amplitudes
of the formation of resonances $a_ie^{i \delta_i}$ and the
amplitudes $\lambda_{ab}$ that, within the framework of the isobar
model, keep in mind the information about three-body interactions
with participation of the spectator pion \cite{FN1}. The detailed
discussion of the crucial approximations of the isobar model can be
found, for example, in Refs. \cite{Re16,As85,Al93,Be11,Be20}. As
noted in Ref. \cite{Aa22}, at present, there are no tools for a
complete description of the amplitudes for thee-body decays from
first principles.
Recently, essential progress in the theoretical description of
three-body decays is associated with dispersion methods, see, for
example, Refs. \cite{Am22,Ro11,Fr14,Ro15,Gi64,Ca06,Ku12,Ku15,Al20,
Al22,Ku22} and references therein. This approach, in principle,
allows one to go beyond the phenomenological isobar model. In
particular, it demonstrates that final-state interactions involving
all three particles in hadronic loops turn out to be important
sources of deviations from the Watson theorem. However, one cannot
but recognize the complexity of applying dispersion methods
\cite{Am22,Ro11,Fr14,Ro15,Ku12,Ku15,Al20,Al22,Ku22} for practical
processing of the data on various three-body decays in comparison
with the isobar model (see especially Refs. \cite{Am22,Ku22}).

The mechanisms of formation of the seed amplitudes
$\lambda_{\pi^+\pi^-}$ and $\lambda_{\pi^0\pi^0}$ in the general
case can differ from each other, as well as the mechanisms of
formation of $\lambda_{K ^+K^-}$ and $\lambda_{K^0\bar K^0}$. If we
take advantage of the language of quark diagrams, then, for example,
due to the $D^+$ decay mechanism indicated in Fig. 2, only a
$K^0\bar K^0$ pair can be produced, while $K^+K^-$ cannot.
Therefore, no isotopic relations between the seed amplitudes of the
different charge state production are assumed in advance.
\begin{figure}  [!ht] 
\begin{center}\includegraphics[width=7cm]{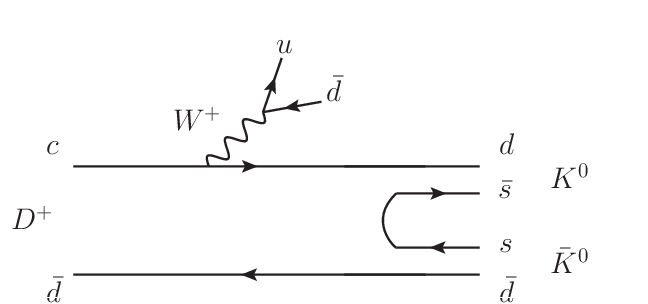}
\caption{\label{Fig2} The tree-level external $W^+$-emission diagram
leading to the $K^0\bar K^0$ pair production in the $D^+$ decay.
}\end{center}\end{figure}

We take the amplitudes $T^0_0(s)$ and $T_{K\bar K\to\pi\pi}(s)=T_{
\pi\pi\to K\bar K}(s)$ from Ref. \cite{Ac06} (corresponding to
fitting variant 1 for parameters from Table 1 therein) containing
the excellent simultaneous descriptions of the phase shifts,
inelasticity, and mass distributions in the reactions $\pi\pi\to\pi
\pi$, $\pi\pi\to K\bar K$, and $\phi\to\pi^0\pi^0 \gamma$ (see also
Refs. \cite{Ac11,Ac12}). The amplitudes $T^0_0(s)$ and $T_{\pi\pi\to
K\bar K}(s)$ were described in Refs. \cite{Ac06,Ac11,Ac12} by the
complex of the mixed $f_0(500)$ and $f_0(980)$ resonances and smooth
background contributions. The amplitude $T^2_0(s)$ is taken from
Ref. \cite{Ac03} (see also Ref. \cite{Ac98a}).


\section{Description of the $D^+\to\pi^-\pi^+\pi^+$ data}


Let us rewrite Eq. (\ref{Eq3}) in terms of the amplitudes $T^0_0(s)
$, $T^2_0(s)$ and $T_{K^+K^-\to\pi^+\pi^-}( s)$ in the following
form
\begin{eqnarray}\label{Eq5}
{\mathcal{A}_{S\scriptsize\mbox{-wave}}(s)=a_0(s)e^{i\delta_0(s)}}=
\lambda_{\pi^+\pi^-}+I_{\pi\pi}(s)\left[T^0_0(s)\left(\frac{2}{3}
\lambda_{\pi^+\pi^-}+\frac{1}{3}\lambda_{\pi^0\pi^0}\right)+T^2_0(s)
\frac{1}{3}\left(\lambda_{\pi^+\pi^-}-\lambda_{\pi^0 \pi^0}\right)
\right]\nonumber \\ +\left[\lambda_{K^+K^-}I_{K^+K^-} (s)+ \lambda_{
K^0\bar K^0}I_{K^0\bar K^0}(s)\right]T_{K^+K^-\to\pi^+ \pi^-}(s).
\qquad\quad\ \ \, \end{eqnarray} Note that if all $\lambda_{ab}$ are
real and $\lambda_{\pi^+\pi^-}=\lambda_{\pi^0\pi^0}$ [i.e., the
contribution of the amplitude $T ^2_0(s)$ is absent], then the
attempt to describe the data \cite{Aa22} about the phase
$\delta_0(s)$ shown in Fig. 3(b) will fail. Really, in this case the
phase $\delta_0(s)$ of the amplitude
$\mathcal{A}_{S\scriptsize\mbox{-wave}}(s)$ [taking into account Eq.
(\ref{Eq4})] coincides with the $\pi\pi$ scattering phase
$\delta^0_0(s)$ below the $K^+K^-$ threshold where $\eta^0_0(s)=1$
[as is the phase of the amplitude $T_{K^+ K^-\to\pi^0\pi^0}(s)$
\cite{Ac06}]. The phase $\delta^0_0(s)$ is shown in Fig. 3(b) by the
dotted curve. We also note that in the vicinity of the $\pi\pi$
threshold, the phase $\delta_0(s)$ is approximately equal to
$100^\circ$ [see Fig. 3(b)], and this cannot be described by any
real constants $\lambda_{ab}$, since the phases $\delta^0_0(s)$ and
$\delta^2_0(s)$ vanish at the $\pi\pi$ threshold and are small in
its vicinity as is seen from Fig. 3(b).

Let us first consider the fitting variant in which the contribution
of the amplitude $T_{K^+K^-\to\pi^+\pi^-}(s)$ is absent, i.e.,
$\lambda_{K^+K^-}=\lambda_{K^0\bar K^0}=0$. In this case, the
connection with the $K\bar K$-channel is taken into account to the
extent that it is present in the amplitude $T^0_0( s)$. This fitting
variant is shown in Fig. 3 by the dashed curves. It corresponds to
the following parameter values:
\begin{eqnarray}\label{Eq6}
\lambda_{\pi^+ \pi^-}=-1.72+i11.30,\ \
\lambda_{\pi^0\pi^0}=17.86+i6.59,\ \ C_{\pi\pi}= 0.77.
\end{eqnarray}
The dash-dotted line in Fig. 3(a) shows the contribution caused by
the amplitude $T^2_0(s)$. Surprisingly, this simple variant quite
satisfactorily describes the observed features of the energy
dependences of the magnitude and phase of the $S$-wave amplitude in
the $D^+\to \pi^-\pi^+ \pi^+$ decay in the region $2m_\pi<m_
{\pi^-\pi^+}<1.39\mbox{ GeV}$.

\begin{figure}  
\begin{center}\includegraphics[width=16.5cm]{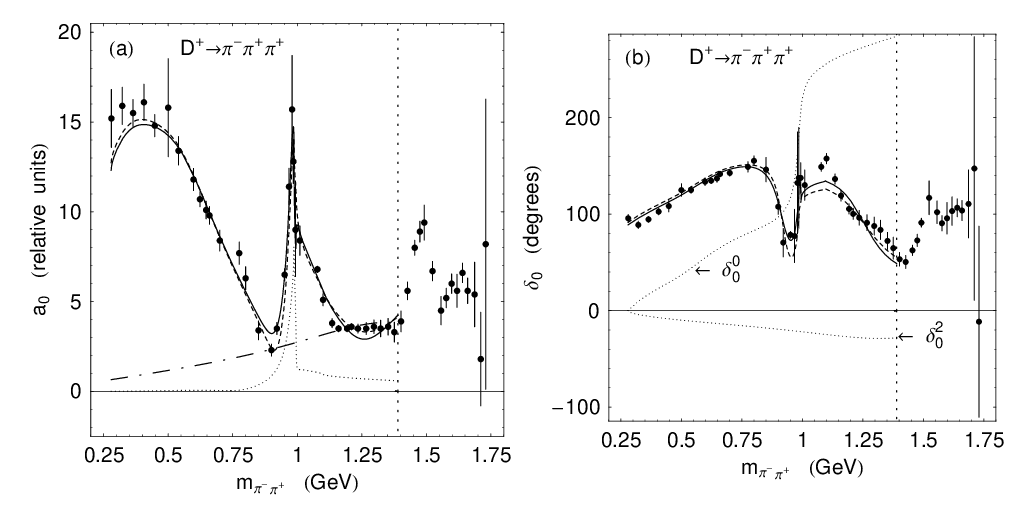}
\caption{\label{Fig3} The points with the error bars are the LHCb
data \cite{Aa22} on the (a) magnitude $a_0$ and (b) phase $\delta_0$
of the $\pi^-\pi^+$ $S$-wave amplitude in the $D^+\to\pi^-\pi^+
\pi^+$ decay. The statistical, experimental systematic, and model
systematic uncertainties are added in quadrature. The solid curves
represent our fit. The corresponding contribution to $a_0$ from the
$T_{K^+ K^-\to\pi^+\pi^-}$ amplitude in Eq. (\ref{Eq5}) is shown in
plot (a) by the dotted curve. The dashed curves show the fit variant
at $\lambda_{K^+K^-}=\lambda_{K^0\bar K^0}=0$. For this variant, the
dash-dot curve in plot (a) shows the contribution from the $T^2_0$
amplitude. The vertical dotted lines show the fitting region
boundary. In plot (b), the dotted curves show the $\pi\pi$
scattering $S$-wave phase shifts $\delta^0_0$ and $\delta^2_0$ which
describe the corresponding data for the reactions $\pi^+\pi^\mp
\to\pi^+\pi^\mp$ well.} \end{center} \end{figure}

The solid curves in Fig. 3 correspond to the fit without any
restrictions on the values of the parameters $\lambda_{ab}$ in Eq.
(\ref{Eq5}) (including $\lambda_{K^+K^-}$ and $\lambda_{K ^0\bar
K^0}$). Formally, this fit (with $\chi^2=162$) turns out to be
noticeably better than the previous variant (with $\chi^2=278$). The
values of the fitting parameters are the following:
\begin{eqnarray}\label{Eq7}
\lambda_{\pi^+ \pi^-}=-1.21+i11.21,\ \
\lambda_{\pi^0\pi^0}=20.40+i4.47,\ \
C_{\pi\pi}=0.68,\ \ \, \nonumber \\
\lambda_{K^+K^-}=39.11+i27.43,\ \ \lambda_{K^0\bar
K^0}=-32.93-i29.98,\ \ C_{K\bar K}=0.46.
\end{eqnarray}
The corresponding contribution to $a_0(s)$ from the amplitude
$T_{K^+K^-\to\pi^+\pi^-}(s)$ is shown in Fig. 3(a) by the dotted
curve. It should be noted that the solid curves and dashed curves
for $a_0(s)$ and $\delta_0 (s)$ presented in Fig. 3 are generally
similar to each other.

Interestingly, the amplitude $a_0(s)$ [module of $\mathcal{A}_{
S\scriptsize\mbox{-wave}}(s)$] reaches its minimum at
$\sqrt{s}=m_{\pi^-\pi^+}\approx0.9$ GeV [see Fig 3(a)], i.e., in the
region where the amplitude of $\pi\pi$-scattering $T^0_0(s)$ reaches
the unitary limit. On the contrary, the $f_0(980)$-resonance
manifests itself in $|T^0_0(s)|$ as a deep and narrow dip, and in
$a_0(s)$ it manifests itself as a resonance peak. By virtue of
chiral symmetry, the resonance $f_0(500)$ (also known as $\sigma$)
is shielded by the background in the $T^0_0(s)$  amplitude
\cite{Ac94,Ac07}. Such a chiral suppression, as can be seen from
Fig. 3(a) is absent in the $a_0(s)$ amplitude. As for the phase
$\delta_0 (s)$, its comparison with the $\pi\pi$ scattering phase
$\delta^0_0 (s)$ [see Fig. 3(b)] explicitly demonstrates a deviation
from Watson's theorem \cite{Wa52}, caused by the difference in the
production mechanisms of the $S$-wave $\pi^-\pi^+$ system in the
$D^+\to\pi^-\pi^+\pi^+$ decay and in $\pi\pi$-scattering.

When describing the peak near 1 GeV in Fig. 3(a), there is no double
counting. Let us extract from the amplitude
$\mathcal{A}_{S\scriptsize\mbox{-wave}}(s)$ in Eq. (\ref{Eq3}) the
contribution with isospin $I=0$ caused by the creation of the
$\pi\pi$ states. In the form suitable below the $K^+K^-$ threshold,
this contribution is
\begin{eqnarray}\label{Eq7a}
\mathcal{A}^0_0(s)=\left(\frac{2}{3}\lambda_{\pi^+\pi^-}+\frac{1}{3}
\lambda_{ \pi^0\pi^0}\right)e^{i\delta^0_0(s)}\left[\cos\delta^0_0
(s)+(\mbox{Re}I_{\pi\pi}(s))\sin\delta^0_0(s)\right].
\end{eqnarray} As paradoxical as it may appear at first
glance, just the dip in the amplitude
$T^0_0(s)=e^{i\delta^0_0(s)}\sin\delta^0_0 (s)/\rho_{\pi \pi}(s)$ in
the $f_0(980)$ region (where the phase $\delta^0_0(s)$ changes very
rapidly and passes through 180$^\circ$) leads to a prominent peak in
the $|\mathcal{A}^0_0(s)|$ near 1 GeV. The contribution of the
$T_{K^+K^-\to\pi^+ \pi^-}(s)$ amplitude [see the dotted curve in
Fig. 3(a)] improves slightly the description of the peak. It is
important to emphasize that these two sources of the peak in the
$a_0(s)$ near 1 GeV have essentially different origins.

To describe the oscillations observed in $a_0(s)$ and $\delta_0(s)$
in the region of $m_{\pi^-\pi^+}>1.39$ GeV (see Fig. 3), additional
considerations are needed about the possible mechanisms production
of the $f_0(137 0)$ and $f_0(1500)$ resonances. Their admixture
(probably small) can enter into
$\mathcal{A}_{S\scriptsize\mbox{-wave}}(s)$ through the
$\pi\pi$-scattering amplitude $T^0_0(s)$. But the $f_0(1370)$ and
$f_0(1500)$, being presumably $q\bar q$-states, may well be directly
produced in the $D^+\to\pi^-\pi^+\pi^+$ decay. In this case, the
corresponding contributions can be described phenomenologically
within the framework of the usual isobar model. In this paper, we do
not dwell on the description of the $m_{\pi^-\pi^+}>1.39$ GeV
region, but we hope to do so elsewhere.


\section{Description of the $D^+_s\to\pi^-\pi^+\pi^+$ data}

Figure 4 shows the LHCb data \cite{Aa22a} for the magnitude $a_0(s)$
and phase $\delta_0(s)$ of the $\pi^-\pi^+$ $S$-wave amplitude in
the $D^+_s\to\pi^-\pi^+\pi^+$ decay. Let us note that the values
given in \cite{Aa22a} for the phase $\delta_0(s)$ are shifted in
Fig. 4 by $+180^\circ$. This is done for the convenience of the
comparison of all three phases $\delta_0(s)$, $\delta^0_0(s)$, and
$\delta^2_0(s)$. The minus sign appearing in Eq. (\ref{Eq3}) as a
result of this shift is absorbed in the coefficients $\lambda_{ab}$.
The solid curves in Fig. 4, which quite successfully describe the
data in the region $2m_\pi<m_{ \pi^-\pi^+}<1.29\mbox{ GeV}$,
correspond to a very simple variant of the model. This variant is
suggested by the very data on the $D^+_s \to\pi^-\pi^+\pi^+$ decay
and by the experience obtained with describing $a_0(s)$ and
$\delta_0(s)$ for the $D^+\to\pi^-\pi^+ \pi^+$ decay. Here we focus
on this variant only.
\begin{figure}  [!ht] 
\begin{center}\includegraphics[width=16.5cm]{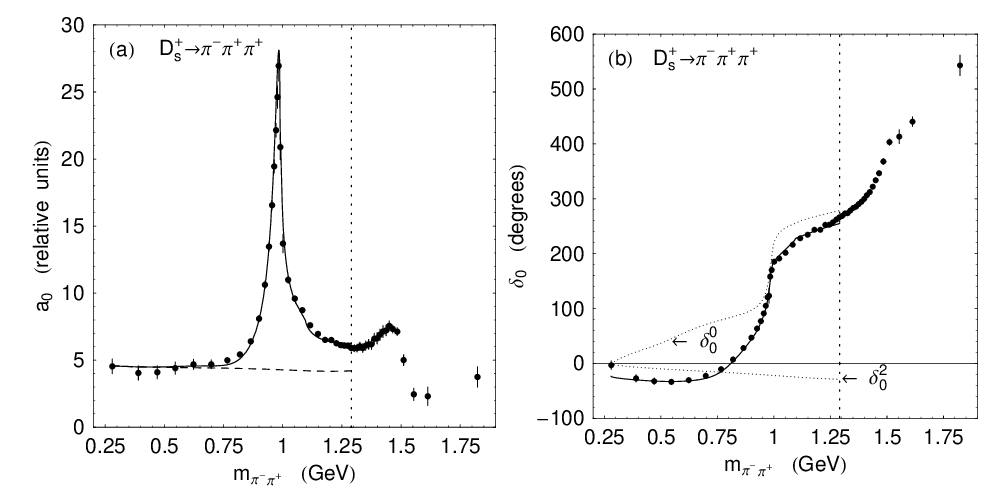}
\caption{\label{Fig4} The points with the error bars are the LHCb
data \cite{Aa22a} on the (a) magnitude $a_0$ and (b) phase
$\delta_0$ of the $\pi^-\pi^+$ $S$-wave amplitude in the $D^+_s\to
\pi^-\pi^+\pi^+$ decay. The statistical, experimental systematic,
and model systematic uncertainties are added in quadrature. The
solid curves represent our fit. The dashed curve in plot (a) shows
the contribution to $a_0$ caused by the term proportional to
$\lambda_{\pi^+\pi^-}$ in Eq. (\ref{Eq8}). The vertical dotted lines
show the fitting region boundary. In plot (b), the dotted curves
show the $\pi\pi$ scattering $S$-wave phase shifts $\delta^0_0$ and
$\delta^2_0$ well describing the corresponding data for the
reactions $\pi^+\pi^\mp\to\pi^+\pi^\mp$.}
\end{center}\end{figure}

When passing from the description of the $D^+$ decay to the
description of the $D^+_s$ decay, we do not change the notations of
the parameters $\lambda_{ab}$ and $C_{ab}$. We put in Eq.
(\ref{Eq5}) $\lambda_{\pi^0\pi^0}=-2\lambda_{\pi^+\pi^-}$ [which
means the suppression of the contribution of the amplitude $T^0_0(s)
$] and $\lambda_{K^+K^-}=\lambda_{K^0\bar K^0}$ [in terms of quark
diagrams, this equality holds, for example, for the seed mechanism
with external radiation of the $W^+$ boson $D^+_c(c\bar s)\to
W^+s\bar s\to\pi^+(K^+K^-+K^0\bar K^0)$]. Thus, we obtain
\begin{eqnarray}\label{Eq8}
{\mathcal{A}_{S\scriptsize\mbox{-wave}}(s)=a_0(s)e^{i\delta_0(s)}}=
\lambda_{\pi^+\pi^-}[1+I_{\pi\pi}(s)T^2_0(s)]+\lambda_{K^+K^-}
\left[I_{K^+K^-} (s)+I_{K^0\bar K^0}(s)\right]T_{K^+K^-\to\pi^+
\pi^-}(s). \end{eqnarray} The solid curves in Fig. 4  demonstrate
the result of the fitting to the data using Eq. (\ref{Eq8}). The
parameter values for this fit (with $\chi^2=129$) are the following:
\begin{eqnarray}\label{Eq9}
\lambda_{\pi^+ \pi^-}=5.37-i2.30,\ \ C_{\pi\pi}=1.69,\ \
\lambda_{K^+K^-}=20.18-i8.94,\ \ C_{K\bar K}=0.60. \end{eqnarray}

In this case, it is almost obvious how each of the contributions
works. In $a_0(s)$ [see Fig. 4(a)], the region of the $f_0(980)$
resonance is dominated by the contribution of the amplitude
$T_{K^+K^-\to\pi^+\pi^-}(s)$. In the region $m_{\pi^-\pi^+}<0.9$
GeV, the contribution of the $f_0(980)$ rapidly falls, and $a_0(s)$
is dominated by the  weakly energy-dependent contribution
proportional to $\lambda_{\pi^+\pi^-}$ in Eq. (\ref{Eq8}). The phase
of this contribution is small, smooth, and negative, like the
$\delta^2_0(s)$ phase [see Fig. 4(b)]. As $m_{\pi^-\pi^+}$
increases, it is compensated due to the rapidly increasing positive
phase of the amplitude $T_{K^+K^-\to\pi^+\pi^-}(s)$ [see Fig. 4(b)],
which, below the $K^+K^-$-threshold, coincides with $\pi\pi
$-scattering phase shift $\delta^0_0(s)$ \cite{Ac06}. As $m_{\pi^-
\pi^+}$ increases further, the description of the $\delta_0(s)$
phase remains quite successful up to $m_{\pi^-\pi^+} \approx1.29$
GeV. About the description of the data in the region of the $f_0(137
0)$ and $f_0(1500)$ resonances, we can only repeat what has been
said at the end of the previous section.


\section{Predictions for the $D^+$ and $D^+_s$ decays into $\pi^+\pi^0\pi^0$}

For the $S$-wave amplitude of the $\pi^0\pi^0$-system produced in
the decay $D^+\to\pi^+\pi^0\pi^0$ we have
\begin{eqnarray}\label{Eq10}
{\mathcal{A}_{S\scriptsize\mbox{-wave}}(s)=a_0(s)e^{i\delta_0(s)}}=
\lambda_{\pi^0\pi^0}+I_{\pi\pi}(s)\left[T^0_0(s)\left(\frac{2}{3}
\lambda_{\pi^+\pi^-}+\frac{1}{3}\lambda_{\pi^0\pi^0}\right)-T^2_0(s)
\frac{2}{3}\left(\lambda_{\pi^+\pi^-}-\lambda_{\pi^0 \pi^0}\right)
\right]\nonumber \\ +\left[\lambda_{K^+K^-}I_{K^+K^-} (s)+ \lambda_{
K^0\bar K^0}I_{K^0\bar K^0}(s)\right]T_{K^+K^-\to\pi^0\pi^0}(s),
\qquad\quad\ \ \, \end{eqnarray} where $T_{K^+K^-\to\pi^0\pi^0}(s)=
T_{K^+K^-\to\pi^+\pi^-}(s)$. The curves for $a_0(s)$ and $\delta_0(s
)$ shown in Fig. 5 are obtained using Eq. (\ref{Eq10}) after
substituting into it the parameter values from Eq. (\ref{Eq7}).
\begin{figure}  [!ht] 
\begin{center}\includegraphics[width=13cm]{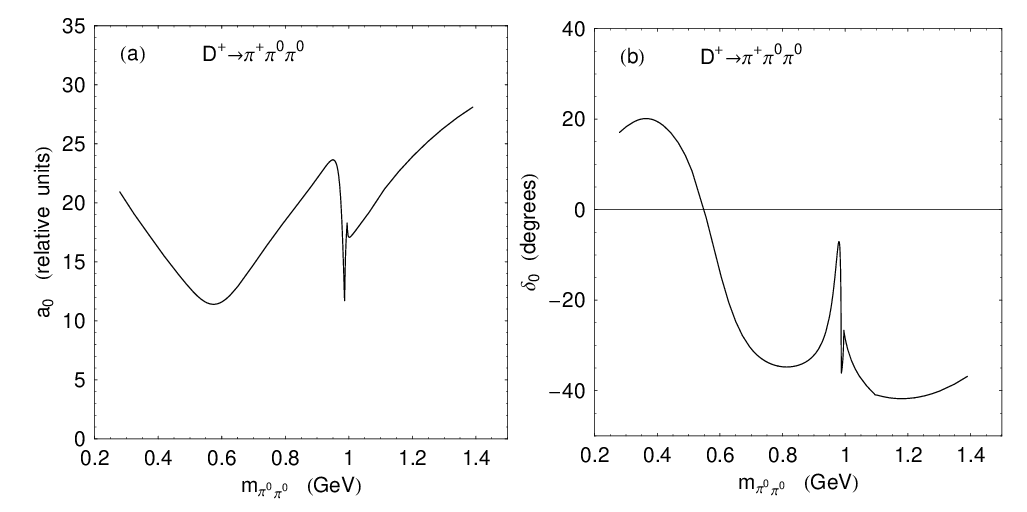}
\caption{\label{Fig5} Predictions for the (a) magnitude $a_0$ and
(b) phase $\delta_0$ of the $\pi^0\pi^0$ $S$-wave amplitude in
$D^+\to \pi^+\pi^0\pi^0$.}\end{center}\end{figure}

An analog of Eq. (\ref{Eq8}) for the $D^+_s\to\pi^+\pi^0\pi^0$ decay
has the form
\begin{eqnarray}\label{Eq11}
{\mathcal{A}_{S\scriptsize\mbox{-wave}}(s)=a_0(s)e^{i\delta_0(s)}}=
\lambda_{\pi^0\pi^0}[1+I_{\pi\pi}(s)T^2_0(s)]+\lambda_{K^+K^-}
\left[I_{K^+K^-} (s)+I_{K^0\bar K^0}(s)\right]T_{K^+K^-\to\pi^0
\pi^0}(s),\end{eqnarray} where $T_{K^+K^-\to\pi^0\pi^0}(s)= T_{K^+
K^-\to\pi^+\pi^-}(s)$ and $\lambda_{\pi^0\pi^0}=-2\lambda_{\pi^+
\pi^-}$. The curves for $a_0(s)$ and $\delta_0(s)$ shown in Fig. 6
are obtained using Eq. (\ref{Eq11}) after substituting into it the
parameter values from Eq. (\ref{Eq9}).
\begin{figure}  [!ht] 
\begin{center}\includegraphics[width=13cm]{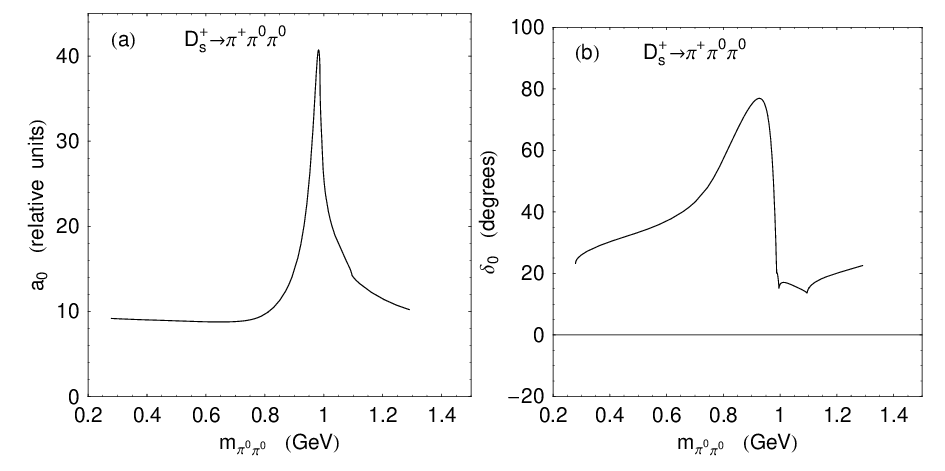}
\caption{\label{Fig6} Predictions for the (a) magnitude $a_0$ and
(b) phase $\delta_0$ of the $\pi^0\pi^0$ $S$-wave amplitude in
$D^+_s\to \pi^+\pi^0\pi^0$.}\end{center}\end{figure}

Comparison of the curves in Figs. 5 and 6 with the corresponding
curves in Figs. 3 and 4 reveals that the predictions obtained for
the decays $D^+\to\pi^+\pi^0\pi^0$ and $D^+_s\to\pi^+\pi^0 \pi^ 0$
are crucial to the verification of the presented phenomenological
model.

\section{Conclusion}

To describe the amplitudes of the $S$-wave three-pion decays of the
$D^+$ and $D^+_s$ mesons, a phenomenological model is presented in
which the production of the light scalar mesons $f_0(500)$ and
$f_0(980)$ occurs due to $\pi\pi$ and $K\bar K$ interactions in the
final state. Such a production mechanism is consistent with the
hypothesis of the four-quark nature of the $f_0(500)$ and $f_0(980)$
states. Using this model, it is possible to satisfactorily describe
virtually all features of the energy dependence of the $\pi^-\pi^+$
$S$-wave amplitudes measured in the $D^+\to\pi^-\pi^+\pi^+$ and
$D^+_s \to\pi^-\pi^+\pi^+$ decays in the regions $2m_\pi<m_{\pi^-
\pi^+}<1.39\mbox{ GeV}$ and $2m_\pi<m_{\pi^-\pi^+} <1.29\mbox{
GeV}$, respectively. The model predictions are presented for the
$D^+\to\pi^+\pi^0\pi^0$ and $D^+_s \to\pi^+\pi^0\pi^0$ decays. Their
verification will be very critical for our model. A problem common
to all isobar models with the explanation of the phases of the meson
pair production amplitudes in multibody weak hadronic decays of
charm states is noted.

The $S$-wave phases measured using the quasimodel-independent
partial wave analysis \cite{Aa22,Aa22a, Ai01,Bo08,Au09,Ab21,Re16}
contain valuable information about the contributions associated with
three-body interactions. But even if the phases of the $ab$
scattering are known, as for the $\pi\pi$ and $K\pi$ systems, to
separate the contributions from the different isospin amplitudes it
is necessary to additionally use a model (for example, of the type
used by us). It can be hoped that for the $ab$ channels with a
definite isospin, the difference between the $S$-wave phase obtained
from $ab$ scattering data and the phase found from the three-body
decay is reduced simply to an overall relative shift, at least in
the elastic region [see, for example, Eq. (\ref{Eq7a})]. For
example, in this way one can determine the phase of $\pi\eta$
scattering up to an additive constant. Thus, it would be very
interesting to perform the quasimodel-independent partial wave
analysis of high-statistics data on the $D^+_s\to\pi^+\pi^0\eta$
decay, in which the $\pi^+\eta$ and $\pi^0\eta$ $S$-wave amplitudes
are parametrized as complex functions determined from the fitting to
the data. It is natural that the found amplitudes can be compared
with theoretical predictions for the elastic $\pi\eta$ scattering.\\

\begin{center} {\bf ACKNOWLEDGMENTS} \end{center}

The work was carried out within the framework of the state contract
of the Sobolev Institute of Mathematics, Project No. FWNF-2022-0021.




\begin{thebibliography}{99}
\bibitem{FN1} The amplitudes $\lambda_{ab}$ [see Eq. (\ref{Eq3}) and Fig. 1] in the isobar
model formally resemble the amplitudes of contact interactions.
However, they are not reduced to the amplitudes of tree point-like
diagrams. These amplitudes are the sources of the production of
non-resonant $ab$ pairs. The amplitudes $\lambda_{ab}$ are
intricately formed complex quantities including, among other things,
projections of the production amplitudes of the resonances in the
crossed channels onto the $S$-wave channels $ab$ and also inelastic
sources of the production of $ab$ states.
\end{thebibliography}
\end{document}